\documentclass{aa}
\usepackage{graphicx}
\begin{document}

\title{Distance and mass of pulsating stars from multicolour \\ 
       photometry and atmospheric models}
\author{S. Barcza\thanks{\email{barcza@konkoly.hu}}}
\institute{Konkoly Observatory, PO Box 67,
                   1525 Budapest, XII,
                   Hungary}
\date{Received 15 October 2002/ Accepted 27 February 2003}

\abstract{
For determining distance and mass of pulsating stars a new, purely 
photometric method is described in which radial velocity observations 
are not needed. From multicolour photometry the variation of angular 
diameter is determined in a conventional way by using the surface 
brightness of the theoretical atmospheric models ATLAS of Kurucz 
(\cite{kuru1}). As a function of phase the following two parameters
are introduced in the Navier-Stokes equation:
gravity (of the appropriate static ATLAS model), and angular diameter.
Distance and mass 
are derived from phases of standstill. Conclusions are drawn on the
hydrodynamic behaviour of the atmosphere. 
The new method is compared with the Baade-Wesselink (BW) method. 
As an example the RR Lyrae variable \object{SU Dra} is given: using 
$UBVR_{\rm C}$
photometry, a distance of
$647\pm 16$pc, 
and a mass of
$0.66\pm .03{\cal M}_{\sun}$ 
were found. In addition to the radius change 
$(4.58-5.51)R_{\sun}$ 
radius  undulations have been found with amplitude 
$\approx 0.2R_{\sun}$
and period 
$P/5$, which are synchronized with the main period
$P=0\fd 66042$.
The present method has confirmed the distance value from the BW method;
thus, in the problem of different distance scales from BW and other methods 
the short extragalactic distance scale has been bolstered by an independent 
argument. The mass is some 30 percent larger than the value from
the equation of pulsation of van Albada \& Baker (\cite{vana1}).
}
\maketitle

\section{Introduction}

The availability of ATLAS stellar atmospheric models (Kurucz,
\cite{kuru1}) made it possible to account for the monochromatic flux
of non-variable stars with a formerly unprecedented
accuracy. As a function of the parameters effective temperature
$T_e$, surface gravity $g$, chemical composition, etc, large grids 
of models are now available for different physical
assumptions on convection and turbulence. 
Ab initio knowledge or reconstruction of photometric filter 
functions made it possible to compute absolute fluxes in a variety of
photometric bands and to fit them to stellar magnitudes and colours by
an appropriate choice of zero points.
The comparison of observed colours of stars with computed colours
of model atmospheres made it possible to determine the
physical parameters of a star. These new possibilities opened
the way to explore the atmospheric parameters of variable stars which
pulsate in such a manner that quasi-static approximation (QSA, Buonaura
et al. \cite{buona}) is valid.

The synthetic colours 
$UBVRIJK$
of new and more sophisticated versions of the
ATLAS models with increased microturbulence were used by Liu \& Janes 
(\cite{liuj1}, \cite{liuj2}) to determine fundamental parameters of RR
Lyrae stars using the Baade-Wesselink (BW) technique. 
There are, however, many problems with the BW method: a theoretical
overview of these was given by Gautschy (\cite{gaut1}) while practical 
questions and the application for RR Lyrae stars were discussed in detail 
by e.g. Liu \& Janes (\cite{liuj2}), Jones et al. (\cite{jone1}).

The present paper will approach the determination of distance and
mass of pulsating stars, especially RR Lyrae stars,
from a modified point of view compared to a conventional the BW analysis.
The synthetic colours and fluxes in different photometric bands of
ATLAS models (Castelli, \cite{cast1}) will be used to determine the
variation of stellar angular diameter by a similar, slightly extended
method as in the BW analysis. However, the angular diameters
and values of the effective gravity 
$g_e$
will be introduced in the Navier-Stokes (N-S) equation averaged over 
the continuum-forming layers: the distance and mass of the star 
will be determined from the momentum balance of the variable atmosphere . 
In this manner the most time consuming radial velocity observations 
are not necessary, their problematic conversion to radius changes 
becomes unnecessary, and the fundamental parameters distance and mass
will be obtained simultaneously by a procedure which is different from 
the BW technique.  

Section 2 describes the method generally and its adaptation to
$UBVR_{\rm C}$ 
photometry which is a compromise between the number of
available observations and the efficiency in determining physical
parameters of a pulsating stellar atmosphere. 
Sect. 3 summarizes the light and colour
curves of \object{SU Dra} (=\object{BD $+67^\circ 652$}) from 
$UBV(RI)_{\rm C}$ 
observations and the results from the method as an example. 
In Sect. 4 the results are discussed and compared with those from
BW analysis. The conclusions are given in Sect. 5.

Spherical symmetry of the pulsation (excitation of radial modes) will 
be assumed. Cgs units will be used
throughout; departures from them will be indicated. 
$r,\theta,\phi$
will denote spherical coordinates with their origin in the stellar
centre of mass.

\section{The method}

Four steps are involved.
\begin{itemize}
\item[1.] The observed magnitudes are converted to fluxes at zero air
mass and corrected for interstellar extinction. 
\item[2.] The observed colour indices are converted to physical
quantities of the theoretical atmospheric models: i.e. to 
$T_e$, $g_e$,
and surface flux in physical units for the different photometric bands 
including bolometric flux. In this step the bolometric flux is obtained from 
$T_e$ 
by the Stefan-Boltzmann law 
while in the photometric bands the zero point of the magnitude scale 
must be determined to obtain the flux in physical units.  
\item[3.] From the combination of observed and model
fluxes the variable angular diameter is obtained
which is introduced in the N-S equation together with the values
$g_e$. 
\item[4.] At a fixed phase (time interval) the N-S equation is an
algebraic equation for the unknown quantities distance and mass. It
must be written out for two phases at least, from which the unknowns
are obtained by elementary operations. 
\end{itemize}

\subsection{Selection of the photometric system}

A photometric system with colour indices
$C_i$ 
is appropriate for the present method
if the physical quantities are single-valued functions 
$T_e(C_1,C_2)$,
${\rm lg}g(C_1,C_2)$,...
and they change slowly, i.e. the finite differences
$\vert\Delta T_e/\Delta C_i\vert$,
$\vert\Delta g/\Delta C_i\vert$,
$\vert\Delta\:{\rm surface\: brightness}/\Delta C_i\vert$
are small for one of the indices
$C_i$
at least. E.g. in the ranges
$6000<T_e<8000$K,
$2<{\rm lg}\:g_e<4$
of an RR Lyrae star for  
$UBV$ 
photometry
\begin{eqnarray}\label{2.1.4}
&& \Delta T_e/\Delta (B-V)\vert_{U-B={\rm const}}
\approx -10^4{\rm K/mag},  \nonumber                 \\
&& \Delta {\rm lg}\:g_e/\Delta (U-B)\vert_{B-V={\rm const}}
\approx -10{\rm dex/mag} 
\end{eqnarray}
which result in errors 
$\approx 100$ K, $0.1$dex
for a typical error
$0.01$ mag
in the colour index. For 
$BVR_{\rm C}$
photometry
\begin{eqnarray}\label{2.1.5}
&& \Delta {\rm lg}\:g_e/\Delta (B-V)\vert_{V-R_{\rm C}={\rm const}}
\approx 70{\rm dex/mag}, \nonumber                 \\
&& \Delta {\rm lg}\:g_e/\Delta (V-R_{\rm C})\vert_{B-V={\rm const}}
\approx -70{\rm dex/mag}
\end{eqnarray}
which make this photometry unfit to determine 
$g_e$.
The final result is that  
$T_e$,
${\rm lg}\:g_e$
and the flux 
${\cal F}$
in different photometric bands are found with reasonable accuracy 
from a pair of the colour indices. The conversion 
$C_i\rightarrow T_e$
etc can be made conveniently using the Kurucz 
tables (\cite{kuru1}) in which the interstellar reddening is included. 

\subsection{The photometric half angular diameter}

We list some useful relations for convenience. At a phase 
$\varphi$ 
in the photometric band 
$x$ 
the luminosity of the star is
\begin{equation}\label{2.1.3}
L_x(\varphi)=4\pi R_0^2(\varphi){\cal F}_x(\varphi)
\end{equation}
where the radius  
$R_0$ 
belongs to zero optical depth
in the reference frame with origin in the stellar centre of mass, 
\begin{equation}\label{2.1.1}
{\cal F}_x=\int_0^\infty{\rm d}\lambda S_x(\lambda)
              \int_0^\infty{\rm d}\tau_\lambda 
              \sum_{i=1}^6c_iB_\lambda(\tau_{\lambda_i})
\end{equation} 
is the physical flux (Kurucz \cite{kuru2}) on the stellar surface
in the photometric band 
$x$
where 
$S_x(\lambda)$ 
is the filter function defining the colour system.
The monochromatic flux on the stellar surface was approximated in
Eq. (\ref{2.1.1}) by a sum,
where 
$c_1,...,c_6$ 
and 
$\tau_{\lambda_1},...,\tau_{\lambda_6}$
are
0.1615, 0.1346, 0.2973, 0.1872, 0.1906, 0.0288,
and
0.038, 0.154, 0.335, 0.793, 1.476, 3.89
respectively (Traving et al. \cite{trav1}),
$B_\lambda$ 
is the Planck function which accounts well for the source function
if the stellar continuum is considered. The meaning of 
Eq. (\ref{2.1.1}) is
that the monochromatic flux originates roughly from six layers between
$R_{\tau_\lambda\approx 3.89} \leq r \leq R_{\tau_\lambda\approx 0.038}$
and the main contribution comes from the neighbourhood of
$R_{\tau_\lambda\approx 0.335}$.
These (wavelength dependent) radii are defined e.g. by
\begin{equation}\label{2.1.6}
\tau_\lambda = 0.335=-\int_{R_0}^{R_{{\tau_\lambda} = 0.335}}
\kappa_\lambda (r){\rm d}r,
\end{equation}
$\kappa_\lambda$
is the monochromatic absorption coefficient. If
$\kappa_\lambda(r)$
is a smooth and moderately changing function of
$\lambda$
we can assume that 
$R_1 = \overline{R_{{\tau_\lambda} = 0.038}}$,
$R = \overline{R_{{\tau_\lambda} = 0.335}}$,
$R_2 = \overline{R_{{\tau_\lambda} = 3.89}}$
are independent of wavelength (overlining denotes averaging over 
$\lambda$).
If
$x$
represents bolometric or an optical band the Rosseland optical depth can be
substituted for
$\tau_\lambda$
to estimate 
$R_1,\: R,\: R_2$,
which are now automatically independent of the wavelength.
$R_0$
is the top of the atmosphere, 
$R_1$, $R_2$
can be regarded as the top and bottom of the continuum-forming 
atmospheric region. 
$(R_1-R_2)/R_0$,
$R/R_0$
vary with 
$\varphi$,
$R_2< R < R_1$.
These variations reflect the change of
${\rm lg}\:g_e(\varphi)$.

At distance
$d$
the observed stellar flux
${\cal I}_x$
will be
\begin{equation}\label{2.2.1}
{\cal I}_x(\varphi)=10^{-A_x/2.5}L_x(\varphi)/4\pi d^2
\end{equation}
where 
$A_x$
is the interstellar extinction in magnitudes. 
The half angular diameter of the star is obtained for zero
optical depth by
\begin{equation}\label{2.2.2}
\vartheta_0(\varphi)=R_0(\varphi)/d
                  =[10^{A_x/2.5}{\cal I}_x(\varphi)/{\cal F}_x(\varphi)]^{1/2}.
\end{equation}
${\cal I}_x$
must be derived from observed magnitudes and compared with the 
tabulated value
${\cal F}_x$
of the theoretical models which are found by a comparison of
observed colour indices with those of computed models. 

The conventional way to determine 
$\vartheta_0(\varphi)$
is to introduce
$A_x$, 
${\cal I}_x(\varphi)=S_010^{-[m_x(\varphi)+{\rm BC}_x(\varphi)]/2.5}$,
and
${\cal F}_{\rm bol}(\varphi)=aT_e^4(\varphi)$
in Eq. (\ref{2.2.2}) where 
$a$
is the Stefan-Boltzmann constant and the zero points
of the magnitude scales
$m_x$,
${\rm BC}_x$ 
are merged in the constant
$S_0$ 
(e.g. Liu \& Janes \cite{liuj2},
$x=V$),
i.e. two parameters
of theoretical models
(${\rm BC}_V$, $T_e$)
belonging to a pair of observed colour indices
are needed. This procedure was 
applied in the present paper as well; the actual form was
\begin{equation}\label{2.1.7}
\vartheta_0=0.636\times 10^{-(V+{\rm BC}_V-2A_V)/5}T_e^{-2}
\end{equation}
in radians. To check Eq. (\ref{2.1.7}) a single parameter
determination of
$\vartheta_0$
was used 
by the following formulae in Eq. (\ref{2.2.2}). An apparent magnitude
$X$ 
is equivalent to a physical flux
\begin{equation}\label{2.2.3}
{\cal I}_X=10^{-7+(X_{\rm Vega}-X)/2.5}\iota_X^{\rm (Vega)}
           {\rm ergs\:sec^{-1}\:cm^{-2}},
\end{equation}
where 
$X_{\rm Vega}=0.03,0.039$ mag,
$\iota_X^{\rm (Vega)}=1.82,1.04$
if
$X=V,R_{\rm C}$,
respectively. In the band 
$X$
the physical flux of a model is
\begin{equation}\label{2.2.4}
{\cal F}_X=10^{-(X_{\rm K}-4.495)/2.5}{\rm ergs\:sec^{-1}\:cm^{-2}}
\end{equation}
where 
$X_{\rm K}$
is the tabulated magnitude of the Kurucz (\cite{kuru1}) tables for
$U,B,V,R_{\rm C},I_{\rm C}$.
The derivation of 
Eqs. (\ref{2.2.3}), (\ref{2.2.4}) is given in Appendix A.

\subsection{The momentum balance of a radially pulsating atmosphere}

For a non-rotating star the 
$r$
component of the N-S equation is
\begin{eqnarray}\label{2.3.1}
& & {{\partial v_r(r,\theta,\phi,t)}\over{\partial t}} 
+v_r{{\partial v_r}\over{\partial r}}
+a_1(r,\theta,\phi,t)                      \nonumber    \\
& & =-{{G{\cal M}}\over{r^2}} 
-{1\over{\rho(r,\theta,\phi,t)}}
{{\partial p(r,\theta,\phi,t)}\over{\partial r}}
+a_2(r,\theta,\phi,t)       
\end{eqnarray}
(Landau \& Lifshitz \cite{lali1})
where
$v_r$
is the 
$r$
component of the velocity,
$G$
is the gravitational constant,
${\cal M}$
is the stellar mass,
$\rho$
is the density.
$a_1=[v_\theta(r,\theta,\phi,t)\partial v_r/\partial\theta+
{\rm sin}^{-1}\theta v_\phi(r,\theta,\phi,t)\partial v_r/\partial\phi
-v_\theta^2(r,\theta,\phi,t)
-v_\phi^2(r,\theta,\phi,t)]/r$
is the rest of the terms from
$(\vec{v,\nabla})v_r$. 
The sum of the terms from the molecular viscosity is
$a_2$.
Setting the sound velocity (about 
$10^6$ ${\rm cm/sec}$)
for the components 
$v_\theta$,
$v_\phi$
as an upper limit, the order of magnitude of
$a_1$
will be
$10$ ${\rm cm/sec}^2$;
therefore, except for phases of stationary flow
(i.e. $\partial v_r/\partial t\approx 0$)
this term is negligible in comparison with the explicitly written terms.
Estimations are not available for 
$a_2$, 
however, it can be assumed to be negligible in comparison with the 
other terms, especially at phases of standstill (i.e. 
$v_r\approx 0$).

For a static stratification the acceleration
\begin{equation}\label{2.3.2}
g_e=-{1\over \rho}{{\partial p}\over{\partial r}}
\end{equation}
is constant within few percents in the domain
$R_2 < r < R_1$.
On the right hand side of Eq. (\ref{2.3.1}) the second term can be 
approximated by a constant
$g_e$
which is obtained from the colour-colour diagram.
To take all dynamical corrections into account the term 
$a_d(r,\theta,\phi,t)=a_1+v_r\partial v_r/\partial r+\cdots$
will be introduced;
`$\cdots$'
represents the terms which were not listed here, e.g. a dynamical
correction in Eq. (\ref{2.3.2}). However, the spatial average 
$g_d$
of
$a_d$ 
is expected to be small in the phases of standstill in comparison to 
the acceleration
$\partial v_r/\partial t$. 

Perfect spherical symmetry of the velocity field will be assumed; the 
terms describing eventual non-radial modes are assumed to be zero:
\begin{equation}\label{2.3.3}
v_r(r,t)
=v_r^{(0)}(t)+{{\partial v_r}\over{\partial r}}\Bigl\vert_{r=R}(R-r)+\cdots
={\dot R}(t)+\cdots,
\end{equation}
the homogeneous velocity 
${\dot R}$
is a good first approximation, the next term would account for the 
compression or expansion of the atmosphere, i.e. for a change of
$R_1-R_2$.
By introducing Eqs. (\ref{2.3.2}), (\ref{2.3.3}) in
Eq. (\ref{2.3.1}) and averaging over the interval
$R_2 \leq r \leq R_1$
(i.e. multiplying by
$r^2{\rm sin}\theta$,
dividing by
$4\pi (R_1^3-R_2^3)/3$,
integrating over
$r,\theta,\phi$)
the equation 
\begin{equation}\label{2.3.4}
{{\partial {\overline v_r}(R,t)}\over{\partial t}}
=-{{G{\cal M}}\over{R^2(t)}}
+g_e(t)+g_d(R,t)
\end{equation}
is obtained.

Now all terms are neglected in Eq. (\ref{2.3.3}) except for 
$\dot R$
and an apparent angular change 
$\vartheta_a=R_a/d$
is introduced such that
\begin{equation}\label{2.3.10}
R_0(t)=R(t)+R_a(t)=[\vartheta(t)+\vartheta_a(t)]d,
\end{equation}
$R_a\ll R$
accounts for the angular change by opacity
as a consequence of the variation of
$g_e$.
In a turning point of
$\vartheta_0$,
i.e. at standstill of the atmosphere, 
${\dot\vartheta}_0\approx 0$,
${\dot\vartheta}_a\approx 0$,
$\vert\partial v_r/\partial t\vert\gg
\vert v_r\partial v_r/\partial r\vert$,
thus, Eq. (\ref{2.3.4}) takes the form
\begin{equation}\label{2.3.5}
[{\ddot\vartheta}_0(t)-{\ddot\vartheta}_a(t)]d
=-G{{\cal M}\over{\vartheta^2(t)d^2}}
+g_e(t)+g_d(t)
\end{equation}
and
$g_d$
is expected to be small.

The functions
$\vartheta_0(t)$
and
$g_e(t)$
can be determined from the observed colour indices and brightness,
${\ddot\vartheta}_0$
can be obtained by numerical differentiation.
$R_a$
can be estimated by interpolation from the Kurucz tables (\cite{kuru2});
compared to
$R_0$
it is negligible if
${\rm lg} g_e > 2.5$.
${\ddot R}_a$
can be estimated by numerical differentiation;
it is negligible during the pulsation except for the phases in which
$g_e$
changes strongly. From continuity of
$\vartheta_0(t)$
it follows that there exist two turning points at least:
$t_1$, $t_2$.
Writing Eq. (\ref{2.3.5}) for 
$t=t_1$ and $t_2$,
neglecting
$g_d$,
and solving the two equations leads to:
\begin{equation}\label{2.3.6}
d={{g_e(t_1)+{\ddot R}_a(t_1)
  -[g_e(t_2)+{\ddot R}_a(t_2)][\vartheta(t_2)/\vartheta(t_1)]^2}
  \over{{\ddot\vartheta}_0(t_1)-{\ddot\vartheta}_0(t_2)
        [\vartheta(t_2)/\vartheta(t_1)]^2}},
\end{equation}
\begin{equation}\label{2.3.7}
{\cal M}=[g_e(t)
          -{\ddot\vartheta}(t)d]\vartheta^2(t)d^2/G.
         \end{equation}
Since 
$\vartheta_a$
is a small correction to
$\vartheta_0$
Eqs. (\ref{2.3.6}) and (\ref{2.3.7}) were solved 
while neglecting the apparent change of radius
($R_a=0$, ${\ddot R}_a=0$, i.e. $\vartheta\equiv\vartheta_0$)
and taking into account
$R_a\not=0$, ${\ddot R}_a\not= 0$, $\vartheta_a=R_a/d$.

\section{Distance and mass of \object{SU Dra}, some physical parameters
of its atmosphere}

As an example some parameters of the RR Lyrae variable
\object{SU Dra} will be determined. This star has a stable light curve
without Blazhko effect; however, the analysis of the 
photometric observations left some ambiguity whether secondary
variations of some 10 mmag do exist (Barcza \cite{barc1}, Paper I).
\object{SU Dra} is a metal-deficient Population II star:
$[M]=-1.6{\rm dex}$. 
Its galactic coordinates are 
$l=133.45$,
$b=+48.27$
i.e. it lies high above the disk, the interstellar reddening is small in
that direction:
$E(B-V)=0.015$ mag
(Liu \& Janes \cite{liuj2}). Its photometric behaviour was described 
in Paper I; from this observational material the segments
$k=25-51$
were used to compile an average
$UBV(RI)_{\rm C}$
light curve because these observations were performed using two comparison
stars. To improve the phase coverage a new segment
$k=52$
was added containing 19
$UBV(RI)_{\rm C}$
observations on
${\rm JD}=2452054$
with the Wright camera attached to the 1m Ritchey-Chretien
telescope of the Konkoly Observatory, Table \ref{t2} reports these 
results. Technical details of the
observations and reduction were the same as in segments
$k=42-51$
(Paper I).

\begin{table}
    \caption{$UBV(RI)_{\rm C}$ magnitudes of \object{SU Dra} on
             ${\rm HJD}=2452054$, segment $k=52$. The fraction of
             HJD is given for $V$, the epochs of $B$, $U$, 
             $R_{\rm C}$, $I_{\rm C}$ are obtained by adding 
             $-.0007$, $-.0018$, $.0005$, $.0010$ to the epoch of $V$.}
    \[
      \begin{tabular}{llrrrrr}
        \hline
        \noalign{\smallskip}
        HJD & $\varphi$ & $V$ & $B$ & $U$ & $R_{\rm C}$ & $I_{\rm C}$ \\
        \hline
        \noalign{\smallskip}
.3906	&.114	&9.492	&9.683	&9.717	&9.322	&9.087	\\
.4072	&.139	&9.546	&9.757	&9.770	&9.363	&9.105	\\
.4111	&.145	&9.559	&9.773	&9.789	&9.367	&9.111	\\
.4150	&.151	&9.573	&9.793	&9.806	&9.381	&9.119	\\
.4314	&.176	&9.614	&9.853	&9.861	&9.402	&9.136	\\
.4353	&.182	&9.624	&9.870	&9.875	&9.432	&9.159	\\
.4391	&.187	&9.636	&9.881	&9.885	&9.416	&9.146	\\
.4553	&.212	&9.676	&9.950	&9.960	&9.466	&9.182	\\
.4592	&.218	&9.686	&9.965	&9.968	&9.458	&9.188	\\
.4630	&.224	&9.694	&9.980	&9.980	&9.473	&9.175	\\
.4791	&.248	&9.737	&10.045	&10.046	&9.492	&9.202	\\
.4830	&.254	&9.750	&10.058	&10.052	&9.503	&9.210	\\
.4868	&.260	&9.762	&10.076	&10.063	&9.500	&9.203	\\
.5062	&.289	&9.797	&10.122	&10.109	&9.528	&9.232	\\
.5110	&.296	&9.807	&10.146	&10.136	&9.546	&9.240	\\
.5148	&.302	&9.817	&10.149	&10.147	&9.551	&9.245	\\
.5311	&.327	&9.845	&10.188	&10.178	&9.557	&9.254	\\
.5349	&.332	&9.848	&10.203	&10.199	&9.567	&9.258	\\
.5388	&.338	&9.854	&10.210	&10.207	&9.576	&9.258	\\
        \noalign{\smallskip}
        \hline
        \end{tabular}
        \]
\label{t2}
\end{table}

\subsection{Observed $UBV(RI)_{\rm C}$ light and colour curves} 

\begin{figure}
  \resizebox{\hsize}{!}
  {\includegraphics{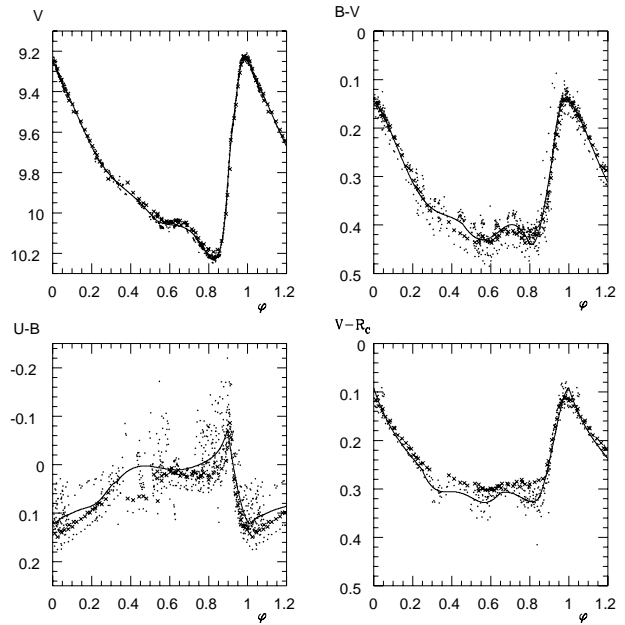}}
  \caption{Light curve $V$ and colour curves of \object{SU Dra}.
           Dots: segments $k=25-29,42-51$ from Paper I and $k=52$,
                 all are shifted by the amount indicated in the text.
           Crosses: segments $k=30-41$ from Liu \& Janes
                 (\cite{liuj1}).
           Lines: fitted curves to segments $k=25-52$.}
  \label{bfig1}
\end{figure}

The observed magnitudes 
$U,B,R_{\rm C}$
were linearly interpolated to the epoch of the 
$V$
observations to avoid systematic errors of colour indices 
$U-B$, $B-V$, $V-R_{\rm C}$
in the rising branch. A comparison with the colour indices of Liu \& Janes 
(\cite{liuj1}) indicated small systematic differences between the colour
systems at Konkoly Observatory and Kitt Peak. Since it is of primary
importance to have the same filter functions in the observations and
computation of synthetic colours of the atmospheric models, the
colours of the Konkoly observations have been shifted by
$\Delta(B-V)=+0.029$,
$\Delta(U-B)=+0.026$,
$\Delta(V-R_{\rm C})=+0.019$ mag.
By these shifts the brightness
$V$
and the colour indices of one of the comparison stars,
\object{BD $+67^\circ 708$}, became identical in
the Konkoly and Kitt Peak colour systems as is obvious from Table 3
of Paper I, and the systematic differences of the Konkoly and Kitt Peak
curves disappeared. Fig.~\ref{bfig1} is a plot of the observed magnitudes
$V$
and the colour indices, the phase was computed according to Paper I
taking into account the secular change of period and phase noise. The
curves drawn in Fig.~\ref{bfig1} were obtained by high-order spline
functions, they were fitted to the shifted colour curves.
These curves are the basis for deriving
$T_e(\varphi)$,
${\rm lg} g_e (\varphi)$,
${\cal F}_X(\varphi)$,
${\cal I}_X(\varphi)$
with
$X=V,R_{\rm C},{\rm BC}_V$,
and
$\vartheta_0(\varphi)$
according to Sect. 2.

The colour curves
$V-I_{\rm C}$
were determined as well; however, they were omitted from the
analysis because the CCD realization of
$I_{\rm C}$
was not free of systematic errors; these originated from the red cut off
which is considerably different from that of a GaAs cathode and of 
the improved  
$S_{I_{\rm C}}(\lambda)$
of Bessell (\cite{bess1}).

The fitted curves show definite structure around
$\varphi\approx 0.4,\: 0.7$
in addition to the ascending branch which is their main feature.
If only 60-70 points are observed (e.g. Liu \& Janes \cite{liuj1})
the fine structure of the light and colour curves remains hidden. 
It could be revealed by the 674 observed points in the 
present study. 

\subsection{Effective temperature, gravity, angular diameter}

\begin{figure}
  \resizebox{\hsize}{!}
  {\includegraphics{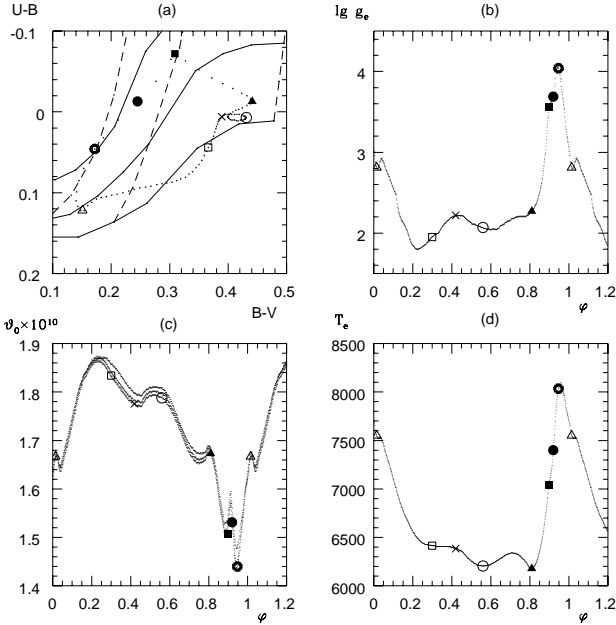}}
  \caption{Panel (a): lines: from bottom to top the isogravity curves
                      ${\rm lg}g=2,3,4$,
                      dashed lines: from left to right the isotherm
                      curves $T_e=8000,7000,6000$ K. The curves are
                      interpolated to $[\mbox{M}]=-1.6$, $E(B-V)=0.015$.
                      Dots: $U-B,B-V$ loop of \object{SU Dra}.
                      Characteristic points of the loop 
                      $\varphi=0.015\: \mbox{and}\: 1.015, 0.3, 0.42, 0.56,
                      0.81, 0.897, 0.92, 0.947$ are marked
                      by different symbols which are plotted in Panels
                      (b)--(d) as well.
           Panels (b), (d): the interpolated ${\rm lg}g_e$, $T_e$ values.
           Panel (c): $\vartheta_0$ from Eq. (\ref{2.2.2}), Eq. (\ref{2.2.3}),
                      Eq. (\ref{2.2.4}) with $V$, $R_{\rm C}$ and from
                      Eq. (\ref{2.1.7}). The symbols of the loop indicate
                      $\vartheta_0(\varphi)$ from Eq. (\ref{2.1.7}).
           }
  \label{bfig2}
\end{figure}

The synthetic colours from the Kurucz (\cite{kuru1}) tables 
(Castelli \cite{cast1}) were interpolated linearly to 
$[M]=-1.6$ dex
and
$E(B-V)=0.015$.
$T_e(\varphi)$, ${\rm lg}g_e(\varphi)$,
tabular magnitudes 
$V$, $R_{\rm C}$, ${\rm BC}_V$
were interpolated to a pair of colour indices
$(U-B,B-V),\: (U-B,V-R_{\rm C}),\: (U-V,V-R_{\rm C})$.
$V$, $R_{\rm C}$ 
were converted to surface flux 
${\cal F}_V$, ${\cal F}_{R_{\rm C}}$ 
by Eq. (\ref{2.2.4}).

Panel (a) of Fig.~\ref{bfig2} shows the theoretical isogravity curves
${\rm lg}g(U-B,B-V)=2,3,4$
for different values of
$T_e$,
the isotherm curves
$T_e(U-B,B-V)=6,7,8\times 10^3$K
for different values of
${\rm lg}g$,
and the observed colour loop
$(U-B,B-V)$
as a function of
$\varphi$. 
Panels (b)-(d) are a plot of
${\rm lg}g_e(\varphi)$, $\vartheta_0(\varphi)$, $T_e(\varphi)$
for the colour loop.
$\vartheta_0(\varphi)$
was determined from Eq. (\ref{2.1.7}) and from
$({\cal I}_V,{\cal F}_V)$, $({\cal I}_{R_{\rm C}},{\cal F}_{R_{\rm C}})$
by Eqs. (\ref{2.2.2}), (\ref{2.2.3}), (\ref{2.2.4}):
it is remarkable that the curves are congruent within 2 percent. This
finding demonstrates that in the range
${\rm lg}g_e$, $T_e$
of an RR Lyrae star the simultaneous use of
$B-V$
and
$U-B$
estimates
$T_e$
well. Therefore, the infrared photometry can be dropped; it is
more complicated from an observational point of view and
was expected to give
$T_e$
correctly by observing the slope of the
continuum over a larger wavelength interval (Liu \& Janes \cite{liuj2}).

\begin{figure}
  \resizebox{\hsize}{!}
  {\includegraphics{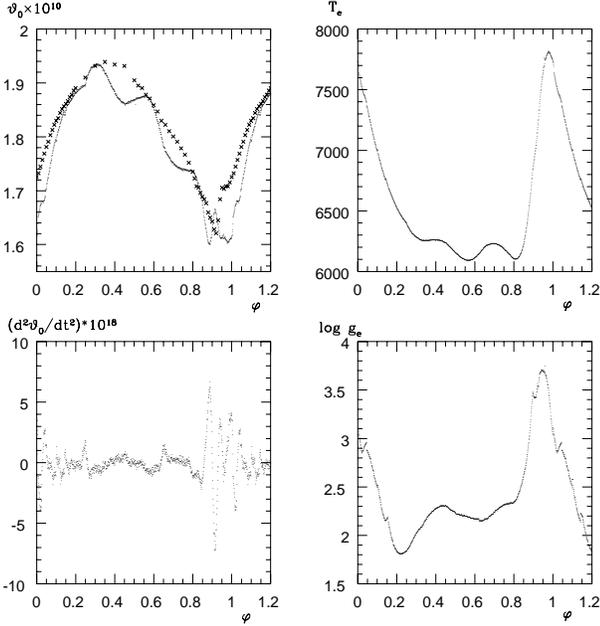}}
  \caption{Dots: averaged curves $\vartheta_0(\varphi)$,
                 $T_e(\varphi)$, ${\rm lg}g_e(\varphi)$ from the loops 
                 $(U-B,B-V),\: (U-B,V-R_{\rm C}),\: (U-V,V-R_{\rm C})$.
                 ${\ddot\vartheta}_0$ was derived from the averaged
                 $\vartheta_0(\varphi)$ by numerical differentiation. 
           Crosses: variation of $\vartheta_0(\varphi)$ given by Liu
                 \& Janes (\cite{liuj2}) from infrared photometry
                 using a formula of type Eq. (\ref{2.1.7}).   
           }
  \label{bfig3}
\end{figure}

The procedure was performed with the colour-colour curves
$(U-B,V-R_{\rm C}),\: (U-V,V-R_{\rm C})$.
The congruence of the curves from Eqs. (\ref{2.1.7}) and (\ref{2.2.2}), 
(\ref{2.2.3}), (\ref{2.2.4}) was within 2 percent confirming that
$T_e$ 
can be determined well by simultaneous use of
$U-B$
or
$U-V$
and
$V-R_{\rm C}$.
The curves 
${\rm lg}g_e(\varphi)$ etc.
were of similar form with those from
$(U-B,B-V)$,
the differences did not exceed 0.1dex, 100 K in the descending branch
while in the ascending branch upper limits of 0.4dex, 
200 K were found for the differences. The curves
$\vartheta_0(\varphi)$
were coincident to within 10 percent. The differences were attributed to
non-perfect filter functions, and to the fitting procedure when the
averaged colour curves were constructed. Among
$\vartheta_0(\varphi)$
from
$(U-B,B-V)$ 
and 
$(U-B,V-R_{\rm C})$, $(U-V,V-R_{\rm C})$
the largest differences were in the phase interval
$0.87<\varphi<1$;
however, the cause for this seems to be that the curves based on
$B-V$
and
$V-R_{\rm C}$
respectively are asynchronous. The differences could have been removed 
by a phase shift; nevertheless, averaging seemed more appropriate. 
Therefore, at the points
$\varphi$
the values
${\rm lg}g_e(\varphi)$, $T_e(\varphi)$, $\vartheta_0(\varphi)$
were averaged from the three combinations of colour-colour indices of 
$UBVR_{\rm C}$. 
The result is plotted in Fig.~\ref{bfig3}.

The values of 
${\rm lg}g_e(\varphi)$, $T_e(\varphi)$
agree well with those Siegel (\cite{sieg1}) and Liu \& Janes (\cite{liuj2}) 
determined from Str\"omgren and
$UBVRIJK$
photometry respectively. The averaged values are
$\overline{{\rm lg}g_e(\varphi)}=2.41$
and
$\overline{T_e(\varphi)}=6490$ K
(this study) while  
$2.69$,
$6400$ K,
and
$2.72$,
$6433$ K
are given by Siegel (\cite{sieg1}) and Liu \& Janes (\cite{liuj2}) 
respectively.

The agreement between the present 
$\vartheta_0(\varphi)$ 
curves and those of Liu \& Janes (\cite{liuj2}) 
is satisfactory. The main difference lies in the undulation of 
the present curve amounting to some 5 percent which
was provided by ten times more photometric observations.
The wavy structure of
$\vartheta_0(\varphi)$
is remarkable: at about
$\varphi\approx 0.45,\: 0.7$
local minima of
$\vartheta_0(\varphi)$
are clearly visible.
This suggests that a sub-oscillation exists which is synchronized with the
main pulsation. In the
phase intervals
$0.8 < \varphi < 0.9$, $0.1 < \varphi < 0.4$
the change of
$\vartheta_a$
and eventual dynamical effects from the strong shock
hamper us in getting a clear picture of the sub-oscillation; 
however, a periodic undulation with
$\approx P/5$
is clearly visible in
$\vartheta_0(\varphi)$
and
$V(\varphi)$
where
$P=5.71\times 10^4$ sec
is the period of \object{SU Dra}. To check the reality of the undulations, 
the 674 
$V$
observations were analysed by the string length
minimization described in Paper I since they can be regarded as the
most reliable because of the use of two comparison stars. The period
$P/5$
has been found only with a very shallow minimum of the string length.

The averaged curve
$\vartheta_0(\varphi)$
was differentiated by midpoint formulae with a phase span of
$\Delta\varphi=0.02(\approx 20$ minutes)
which was found to be an appropriate compromise between the noise and 
the accuracy of
${\ddot\vartheta}_0$.
This choice is in agreement with the present simplified hydrodynamic
model: a description of more rapid structures cannot be expected. 
The angular acceleration
${\ddot\vartheta}_0$
is plotted in the lower left panel of Fig.~\ref{bfig3}. 

\begin{figure}
  \resizebox{\hsize}{!}
  {\includegraphics{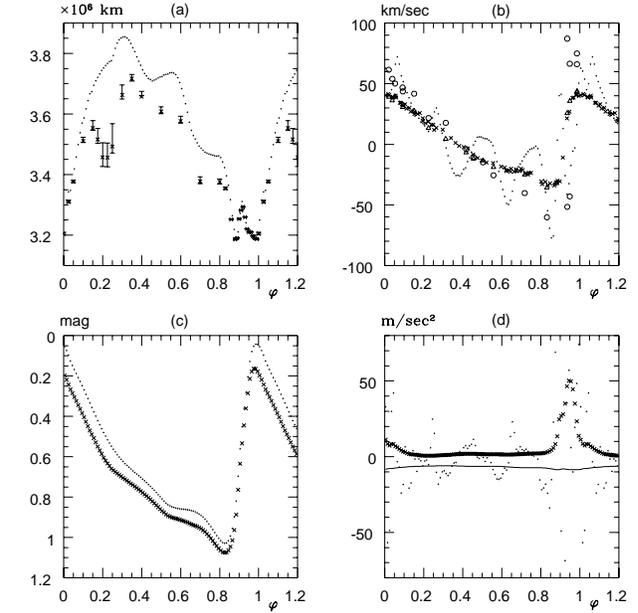}}
  \caption{Physical quantities characterizing the pulsation of
           \object{SU Dra}, $d=1.99\times 10^{16}$ km was used to
           convert the angular quantities.
           Panel (a): dots: $R_0(\varphi)$,
                      crosses: $R(\varphi)$
                      which belong to 
                      $\tau=0$ and  
                      $\tau_{\rm Ross}=0.335$ respectively.
                      The error bars indicate 
                      $R_1(\varphi)-R_2(\varphi)$.
           Panel (b): dots: ${\dot\vartheta}_0d$,
                      crosses: from CORAVEL radial velocities (Liu \& 
                               Janes \cite{liuj1}),
                      triangles and circles: 
                      from radial velocities of metallic lines and 
                      ${\rm H}_\gamma$        
                      (Oke et al. \cite{oke1})
                      $1.32(-v_{\rm radial}(\varphi)-166.9)$ km/sec is plotted.
           Panel (c): dots: $V-9.2$
                      crosses: $M_{\rm bol}=
                                -2.5{\rm lg}[L(\varphi)/L_{\sun}]+4.75$.
           Panel (d): dots: ${\ddot\vartheta}_0d$,
                      line: $-G{\cal M}/R^2(\varphi)$,
                      crosses: $g_e$.    
           }
  \label{bfig4}
\end{figure}

To estimate the apparent radius change
$R_a(\varphi)$
the geometrical depth of 
$\tau_{\rm Ross}=0.335$ 
was interpolated to
${\rm lg}g_e(\varphi)$, $T_e(\varphi)$
from the Kurucz (\cite{kuru2}) tables. 
$R_0(\varphi)=1.99\times 10^{16}\vartheta_0(\varphi)$ km,
$R(\varphi)$,
and 
$R_1(\varphi)-R_2(\varphi)$
as error bars are plotted in Fig.~\ref{bfig4}a.
From the point of view of apparent diameter change
and  acceleration the critical phase intervals are
$0.8 < \varphi < 0.9$,
$0.1 < \varphi < 0.4$.
The break of the slope of
$\vartheta_0(\varphi)$
at
$\varphi\approx 0.25$
can be interpreted by the peak of
$R_a$
which is superimposed on the true atmospheric motion 
$R_0(\varphi)$.
${\ddot R}_a$
was obtained from differentiating polynomial fits to
$R_a$.
Three points are noteworthy:
\begin{itemize}
\item[$\ast$] The secondary peak of
$\vartheta_0$
at
$\varphi\approx 0.55$
coincides just with the fall of the infrared brightness
$K$
which begins at phase delay
$\approx 0.6$
from the visual maximum (Liu \& Janes \cite{liuj1}).
\item[$\ast$] Double standstill exists at minimal radius with differences
$\Delta\varphi=0.031$,
$\Delta \vartheta_0=\vartheta_0(0.947)-\vartheta_0(0.978)
\approx 10^{-12}$
(i.e. $\Delta R_0\approx 2\times 10^4$km).
\item[$\ast$] The most prominent features of
$\ddot\vartheta_0$
are produced by a strong variation of 
$R_a=R_0-R$.
\end{itemize}

\subsection{Distance and mass}

At both turning phases the fine structure of
$\vartheta_0(\varphi)$
shows two standstills with roughly equal
$\vartheta_0$,
the averaged values of the quantities for Eqs. (\ref{2.3.6}), (\ref{2.3.7}) 
are listed in Table \ref{t3}. 
Selection criteria were for these phases that the
atmosphere must be in standstill to have the smallest 
$g_d$
in Eq. (\ref{2.3.4}), 
the effective gravity must be constant to avoid strong variation of
$R_a$, 
and the interval of the averaging must be long enough to smooth out 
scatter and eventual
rapid variations for which the present hydrodynamic model is
oversimplified. The errors of
${\rm lg}g_e$
and
$\vartheta_0$
are less than 3 percent; the errors of
${\ddot\vartheta}_0$
and
${\ddot R}_a$
dominate the error of 
$d$,
$\cal M$.
In the phase intervals A and B the atmosphere is falling freely, in C and D
it is in the state of maximal compression by the shock(s?) from the deeper 
layers.  

In the intervals A,B,C,D it can be assumed that the terms of 
$g_d$
with a factor
$v$
are negligible, i.e. 
$g_e$
of the static atmosphere accounts well for the outward acceleration 
via Eq. (\ref{2.3.4}). The values 
$g_e$
etc of the pairs AC, AD, BC, BD, CD were 
introduced in Eqs. (\ref{2.3.6}) and (\ref{2.3.7}), 
$d$
and
$\cal M$
are listed in Table \ref{t4}, columns 2,3 and 4,5 refer to 
$R_a=0$, ${\ddot R}_a=0$
and 
$R_a$, ${\ddot R}_a$ 
from Table \ref{t3}, respectively. The pair AB was omitted because the 
errors in
$g_e$, ${\ddot\vartheta}_0$, ${\ddot R}_a$
made
$d$
and
$\cal M$
unrealistic. After averaging columns 2,3
$d=(564\pm 31)$pc,
${\cal M}=(0.51\pm .06){\cal M}_{\sun}$
were found as a first approximation; the errors represent 68 percent confidence
level. 

Columns 4,5 are more surprising: they suggest 
\begin{eqnarray}
&&d=(647\pm 16){\rm pc}             \nonumber  \\
&&{\cal M}=(0.66\pm .03){\cal M}_{\sun}. \nonumber
\end{eqnarray}
The values using A were excluded from these averaged 
$d$, ${\cal M}$
on the following arguments. The atmosphere is very extended in the interval
$0.15 < \varphi < 0.4$
i.e. its optical scale height increases to
$R_0-R_2\approx 3\times 10^5{\rm km}\approx 0.1R_0$
at
$\varphi=0.225$,
plane parallel approximation (used in ATLAS models
for radiative transfer) is questionable
(Baschek et al. \cite{basch}).
The apparent acceleration exceeds the effective gravity
(${\ddot R}_a\approx 2g_e$),
$g_d\approx 0$
is presumably not satisfied,
and it begins just a strong apparent (linear rather than accelerating?)
contraction; furthermore,
${\ddot R}_a$ 
can be estimated with large uncertainty. In Table \ref{t3}
${\ddot R}_a(\varphi=0.31)=2.27{\rm m/sec^2}$
is probably an overestimation, e.g. an ad hoc correction to 
${\ddot R}_a(\varphi=0.31)=0.10{\rm m/sec^2}$
would result in
$d=587,559$pc
and
${\cal M}=0.81,0.70{\cal M}_{\sun}$
which are somewhat closer to the values from combining B,C,D.
Therefore, the data from AC, AD have a very low weight; however, 
to give a formal standard error to the unequivocal 
$d$
and 
$\cal M$
from BC,BD,CD they were used but omitted in the averaging. 

\begin{table}
    \caption{The averaged effective gravity, apparent acceleration
             of the continuum-forming layers, angular diameter, and
             acceleration for the phase interval $\varphi$ 
             of standstills. The units are
             ${\rm m/sec^2,\: \rm m/sec^2,\: 10^4km,\: 
              10^{-10}radian,\: 10^{-19}radian/sec^2}$. 
             }
      \[
          \begin{tabular}{lrrrrr}
           \hline
$\varphi$ & $g_e$ & ${\ddot R}_a$ & $R_a$ &  $\vartheta_0$ &
                                                     $\ddot\vartheta_0$ \\
           \hline
            \noalign{\smallskip}
A: .305-.315    &  1.12     & $ 2.27:$ & 16.9 & 1.93  &  $-4.75\pm .56$    \\
B: .545-.555    &  1.58     & $-0.25 $ & 11.3 & 1.88  &  $-2.67\pm .45$    \\
C: .941-.954    &  50.1     & $ 3.45 $ & 0.49 & 1.61  &  $22.6\pm  2.5$    \\
D: .973-.983    &  31.6     & $ 4.02 $ & 0.99 & 1.60  &  $13.6\pm  1.3$    \\ 
            \noalign{\smallskip}
           \hline
          \end{tabular}
      \]
\begin{list}{}{}{}
\item[$:$ uncertain value]
\end{list}
\label{t3}
\end{table}

\begin{table}
    \caption{Distances and masses from Eqs. (\ref{2.3.6}), (\ref{2.3.7}) 
             without and with correction for $R_a$, 
	     in units of pc and ${\cal M}_{\sun}$ respectively.
             }
      \[
          \begin{tabular}{|l|rr|rr|}
           \hline
    & \multicolumn{2}{c|}{I} & 
      \multicolumn{2}{c|}{II}     \\
    &  $d$ & $\cal M$ & $d$ & $\cal M$          \\
           \hline
AC    &  535    & 0.68 & 555 & 0.85              \\
AD    &  476    & 0.49 & 512 & 0.68              \\
BC    &  593    & 0.57 & 647 & 0.66              \\
BD    &  554    & 0.48 & 647 & 0.66              \\
CD    &  665    & 0.31 & 647 & 0.66              \\
           \hline
          \end{tabular}
      \]
\begin{list}{}{}{}
\item[I: $R_a=0,\: {\ddot R}_a=0$]
\item[II: $R_a$ and ${\ddot R}_a$ from Table \ref{t3}, 
          $\vartheta_a=R_a/647$pc]
\end{list}
\label{t4}
\end{table}

\subsection{Luminosity, radius, velocity, acceleration}

The distance modulus is
\begin{eqnarray}
m-M=(9.05\pm .05)\: {\rm mag}.    \nonumber
\end{eqnarray}
Using
$R_0(\varphi)$
and
$T_e(\varphi)$
the intensity averaged luminosity is
$\langle L\rangle=(1.68\pm .08)\times 10^{35}{\rm ergs/sec}
=(42.4\pm 2.1)L_{\sun}$;
the errors were derived from the error of
$d$. 
The mean radius is 
$\overline{R_0}=3.54\times 10^6$km.
The intensity and magnitude averaged visual absolute magnitudes are
$\langle M_V^{\rm int}\rangle=0.74\pm .05$
and
$\langle M_V^{\rm mag}\rangle=0.78\pm .05$
respectively. 

Fig.~\ref{bfig4}b-d are plots of different physical 
quantities derived with
$d=1.99\times 10^{16}$ km,
the data were averaged for intervals
$\Delta\varphi=0.01$.
Note that the velocity
${\dot\vartheta}_0(\varphi)d$
shows a correlation with
$p[v_\gamma-v_{\rm radial}(\varphi)]$
only if it is averaged over a longer time.
($v_\gamma=-166.9$km/sec
(Liu \& Janes \cite{liuj2}) is the barycentric velocity,
$p=1.32$
is the conversion factor.) 
The phase difference between bolometric and visual maximum is 
$\Delta\varphi\approx 0.015$. 

\section{Discussion}

\subsection{Appraisal of the method, comparison with the BW technique}

The surface brightness method to determine stellar angular diameters
is known in two variants. One of them (Barnes et al. \cite{barne})
uses the empirically determined surface brightness of non-variable stars
and applies it to variables. The other variant is bound
to static theoretical atmospheric models which are the sources 
of surface brightness and effective gravity. When
applying them to variable stars a reliable surface brightness can be
determined in the phases where a strong deviation from QSA 
is not expected. In the present paper the theoretical variant 
was used: the half angular diameter and the effective gravity were 
obtained from ATLAS atmospheric models as a function of phase. 
Concerning RR Lyrae stars arguments for the applicability of QSA
were discussed in detail by Buonaura et al. (\cite{buona}). Step 1 is
identical in the present and BW methods. Step 2 
is nearly the same in both methods, the difference is
only that here the values
$\vartheta_0$
from effective temperature and
bolometric correction to 
$V$
are checked by a direct comparison of fluxes in
$V$
and 
$R_{\rm C}$.

Steps 3 and 4 are principally different from the BW
technique: all the data 
$\vartheta_0(\varphi)$, $g_e(\varphi)$
are used in a dynamical equation, i.e. in the 
$r$
component of the N-S equation averaged over the continuum-forming
layers. Some terms could be neglected and the
pressure-density stratification of atmospheric models with QSA
was introduced. An essential assumption is that the function
$g_e=g_e(C_1,...)$
gives the uniform outward acceleration in
$R_1\leq r\leq R_2$,
and the sum 
$g_e$
plus gravity determines the motion of the atmosphere at
$R$.
A term
$g_d$
was defined to account for the dynamical corrections to the 
stratification of QSA. In the range 
$T_e$, 
${\rm lg}g_e$
of RR Lyrae stars the ATLAS models are convective. Its presumably small 
effect on the momentum balance of the pulsation could be incorporated in
$g_d$;
however, it was neglected by the use of Eq. (\ref{2.3.2}).
In standstills of the atmosphere
$g_d$
is expected to be negligible and the averaged 
N-S equation is simplified to algebraic  equations for the unknown 
quantities distance and mass. In the present method the mass plays an 
inherent role; it is determined simultaneously with the distance. 

In the BW method the mass of a pulsating star plays an indirect role: 
in an equation of type Eq. (\ref{2.3.4}) with 
$g_d\equiv 0$
an assumed mass is introduced to derive
$g_e(\varphi)$
by differentiating the curve 
$-pv_{\rm radial}(\varphi)$
numerically.
This effective gravity is then used to select the atmospheric models
from which the effective temperature is determined by a relation of type
$T_e(C_i)$,
and the variation of angular diameter is determined. The kinematic
equation
\begin{eqnarray}\label{4.1.1}
[\vartheta_0(\varphi_2)-\vartheta_0(\varphi_1)]d
      &  = &  R_0(\varphi_2)-R_0(\varphi_1)  \nonumber           \\
      &  = & \int_{\varphi_1}^{\varphi_2}p[v_{\gamma} 
              -v_{\rm radial}(\varphi)]P{\rm d}\varphi
\end{eqnarray}
is solved for
$d$
by various techniques (e.g. Liu \& Janes \cite{liuj2},
Cohen \cite{cohe1}, Jones et al. \cite{jone1}),
in the interval
$[\varphi_1,\varphi_2]$
the atmospheres must be free of shocks. The primary error sources are
$v_{\gamma}$
and
$p$;
it is obvious that their uncertainty can result in a considerable error in
$d$
(Sabbey et al. \cite{sabb1}, Fernley \cite{fern1}). 
The undulations reported in the present paper have a negligible effect on
$d$
if
$\vert\varphi_2-\varphi_1\vert$
is sufficiently large. Another small correction comes from the fact that
$R(\varphi)$
would be more appropriate than
$R_0(\varphi)$
if the radius change is determined by using
radial velocities from CORAVEL (e.g. Liu \& Janes \cite{liuj1}) or 
from weak metallic lines (e.g. Oke et al. \cite{oke1}).

In the present method 
$v_{\rm radial}(\varphi)$, 
$v_{\gamma}$,
$p$,
and an eventual arbitrary phase shift of the radius change from 
photometry and radial velocity (e.g. 
$\Delta\varphi=0.02$
for \object{SU Dra}, Siegel \cite{sieg1})
do not enter at all, the observation of radial velocities is
not needed. However, a colour must observed, e.g. 
$U$
in Johnson system, which is a good indicator for gravity. In the BW method
$g_d\equiv 0$
can result in a negligible error, in the present method 
at selected phases
$g_d=0$, 
i.e. the unconditional use of Eq. (\ref{2.3.2}),
and the simplification of the velocity field to Eq. (\ref{2.3.3}) with a
single constant term can be assumed as a source of more uncertainty
than in the BW method which involves the assumption 
$v_r(r,t)={\dot R}(t)$
as well.
Problems arising from the difference
$S_x(\lambda)-S_x^{\rm o}(\lambda)\not\equiv 0$
were discussed by Bessell (\cite{bess1}) where 
$S_x^{\rm o}(\lambda)$
represents the filter functions of an observatory. They are common both
in the present method and in the BW method. This eventual systematic 
error is phase dependent.

\subsection{The results for \object{SU Dra}}

The distance derived in the present study agrees very well with its value 
$d=640$ pc
from the BW analysis (Liu \& Janes \cite{liuj2}) if the subtle changes of
$R_a(\varphi)$
are taken into account. It is in accordance with the low absolute 
brightness of an RR Lyrae star if it is determined from a direct 
analysis. The higher absolute brightness of an RR Lyrae star from
cluster properties is not bolstered by the present value of
$d$.
Concerning the short and long extragalactic distance scale
(Gratton \cite{grat1}) the discrepancy has been strengthened and
an additional independent argument has been given 
for the short scale. If the change of
$R_a(\varphi)$
is neglected the discrepancy gets worse by
$d=564\pm 31$pc.
Furthermore, Eq. (\ref{4.1.1}) has been
verified: 
$p=1.32$
and
$v_\gamma=-166.9$ km/sec
(Liu \& Janes \cite{liuj2})
must essentially be correct since e.g. for
$\varphi_2=0.8$, $\varphi_1=0.4$
an error
$\Delta v_\gamma=1$ km/sec
would result in an error
\begin{equation}\label{4.2.1}
\Delta d/d=[(\varphi_2-\varphi_1)pP\Delta v_\gamma]
          /[R(\varphi_2)-R(\varphi_1)]\approx 0.04
\end{equation}
i.e. roughly $-0.09$ mag in
$M_V$.
(It must be noted that Oke et al. (\cite{oke1}) argue for
$v_\gamma=-161$ km/sec).
In the BW method the secondary error sources undulation and use of
$R$
for
$R_0$
result in
$\vert\Delta d/d\vert<0.005$.

Liu \& Janes (\cite{liuj2}) assumed 
${\cal M}=0.55-0.65$
to get
${\rm lg} g_e$
for the interval
$0.4\leq\varphi\leq 0.8$;
they remarked that their
$d$
was insensitive to the actual value
($0.6{\cal M}_{\sun}$). 
The basic pulsation
equation of van Albada \& Baker (\cite{vana1}) gives
${\cal M}_{\rm SU\: Dra}=0.47{\cal M}_{\sun}$,
from which the present value
$0.66\pm .03{\cal M}_{\sun}$
differs significantly, however,
its acceptance must be supported by a smaller uncertainty in input physics.

The observed 
$-pv_{\rm radial}(\varphi)$
(Liu \& Janes \cite{liuj1}, Oke et al. \cite{oke1}) does not show the 
undulation of
${\dot\vartheta}_0d$,
and the assumption of an eventually phase dependent conversion factor 
(Sabbey et al \cite{sabb1}) cannot bring the curves to congruence. 
For a few variables in M5 and M92 a similar undulation of
$\vartheta_0(\varphi)$
was shown graphically by Cohen (\cite{cohe1}); however, it was not
discussed. On a speculation 
level the most probable cause of this inconsistency is the non-negligible
velocity gradient coupled with undulation of
$\vartheta_0$.
The radial velocities (especially those from CORAVEL) reflect the averaged 
velocity from the deep layers
$\tau_{\rm Ross}\approx 0.335$ 
while  
${\dot\vartheta}_0d$
is the averaged velocity over the layers
$0\approx\tau<\tau_{\rm Ross}=0.038$. 
At phase
$\varphi\approx 0.75$
a weak hump is discernible on the CORAVEL curve which is correlated with 
the local maximum 
$\vartheta_0(0.78)$.
It is, however, puzzling why there is such a loose correlation of
${\dot\vartheta}_0(\varphi)d$ 
with
$p[v_\gamma-v_{\rm radial}(\varphi)]$.

\section{Conclusions}

In the frame of quasi-static approximation for the atmosphere a new
method has been described to determine distance and mass of pulsating 
stars. In elaborating the photometry there have been minor improvements 
compared to the conventional way in the BW analysis: the use of 
bolometric magnitudes from $T_e$ minus the bolometric correction was
supplemented with a comparison of physical fluxes from models and
observations in a single filter band. It was postulated that 
at the phases of standstill the static ATLAS models give the 
effective outward acceleration in the atmosphere correctly. The
effective gravities,
the half angular diameters and their second derivatives with respect
to time were introduced in the 
$r$ 
component of the Navier-Stokes equation from which we found distance and 
mass while neglecting dynamical corrections in the standstills of
the atmosphere and making a small correction for the apparent 
radius changes. The method presented is purely photometric; radial
velocity observations and their sophisticated interpretation are not needed. 

As an example 
the light and colour curves of \object{SU Dra} have been interpreted.
The agreement of the photometric angular diameters of different
origin was within the expected error of a few
percent. The distance we found is in 
excellent agreement with that from the BW
analysis. A higher mass was found than that from pulsation
equations; however, it is well reconcilable with the general mass values of
RR Lyrae stars. An undulation of angular diameter has been found which
can originate from a surface wave, 
restricted to the outermost parts of the atmosphere. It is synchronized
with the main pulsation.

The results are encouraging for an application to distant RR Lyrae
stars, e.g. for members of globular clusters or nearby galaxies. The
necessary photometric observations are simple and can be done for
large samples of these faint stars by moderate efforts. 
Distance derivations based on the BW method are available 
for some 30 field and a few cluster RR Lyrae variables 
(Gratton et al. \cite{grat2}). These limited numbers could easily be
increased by an order of magnitude if the present method were applied.

\begin{acknowledgements}
The author is grateful to J. M. Benk\H o, Z. Koll\'ath, L. Szabados
for comments on the manuscript,
to Fiorella Castelli for communicating her unpublished colours of Vega 
computed with the revised filters, and to J. Katgert for improving
English presentation.

\end{acknowledgements}

\appendix
\section{The zero point of the magnitudes
$UBV(RI)_{\rm C}$
of the Kurucz tables}

T\"ug et al. (\cite{tugh}) measured the physical flux 
${\cal I}_\lambda^{({\rm Vega})}$
of Vega at
90 wavelength in the interval
$3295\mbox{ \AA} \leq \lambda \leq 9090\mbox{ \AA}$
(Flagstaff flux calibration for $\alpha$ Lyr); this is 
${\cal I}_\lambda$ 
for Eq. (\ref{2.2.2}) in units
${\rm ergs\:sec^{-1}\:cm^{-2}\:\mbox{ \AA}^{-1}}$. 
The theoretical Eddington flux of Kurucz (\cite{kuru1}, 
file {\tt fvega.pck} belonging to
$T_e=9550$K,
${\rm lg}\:g=3.95$,
$[{\rm metallicity}]=-0.5{\rm dex}$,
$v_{\rm turb}=2$km/s) 
was multiplied by
$4\pi$,
converted to the same units and interpolated to the 90 wavelengths:
this is 
${\cal F}_\lambda^{({\rm Vega})}$ 
for Eq. (\ref{2.2.2}). 
$A_\lambda$ 
was set to 
$=0$,
$\vartheta_0^{({\rm Vega})}=(7.95\pm .02)\times 10^{-9}$
was found which agrees well with the interferometric value
$(7.85\pm .17)\times 10^{-9}$
(Hanbury Brown et al. \cite{hanb1}). Having
${\rm BC}_V=-0.310,\: T_e=9550\:{\rm K}$
from the theoretical model and
$V=0.03$ mag
from the observations the constant
$0.636$
is obtained in Eq. (\ref{2.1.7}).
(The parameters differ slightly from the observational values of 
Code et al. (\cite{code1}):
${\rm BC}_V=-0.250,\: T_e=9660\: {\rm K}$.)

The observed flux
${\cal I}_\lambda^{({\rm Vega})}$
and the theoretical flux
$\vartheta_0^{({\rm Vega})}{\cal F}_\lambda^{({\rm Vega})}$
were integrated with the filter functions for the system
$UBV(RI)_{\rm C}$
(Bessell \cite{bess1}) normalized to unity
to obtain the physical fluxes belonging to
the magnitudes of Vega. Table \ref{t1} shows the results.
${\cal I}_U$
could not be computed from the T\"ug et al. (\cite{tugh}) calibration
because of a lack of data in the ultraviolet. The difference of 
${\cal I}_X^{({\rm Vega})}$ 
and 
$\vartheta_0^{({\rm Vega})}{\cal F}_X^{({\rm Vega})}$
originates from lines:
${\cal I}_\lambda^{({\rm Vega})}$
was not measured at the strong lines and Balmer jump; it was  
interpolated linearly.

\begin{table}
    \caption{Magnitudes and physical fluxes of Vega in units 
             $10^{-8}{\rm ergs\:sec^{-1}\:cm^{-2}}$
             }
      \[
          \begin{tabular}{lrrrrr}
           \hline
            \noalign{\smallskip}
$X$& $U$ & $B$ & $V$ & $R_{\rm C}$ & $I_{\rm C}$              \\
           \hline
            \noalign{\smallskip}
magnitudes & $0.022$ & $0.027$ & $0.03$ & $0.039$ & $0.035$\\
${\cal I}_X^{({\rm Vega})}$& ---   &  32.9	&  18.1  &  10.5  &  5.95    \\
$\vartheta_0^{({\rm Vega})}{\cal F}_X^{({\rm Vega})}$&21.6&31.7&18.2&10.4&5.87\\
            \noalign{\smallskip}
           \hline
          \end{tabular}
      \]
\label{t1}
\end{table}

For computing the observed flux of a star at zero air mass
the lower row of Table \ref{t1} must be used, an apparent magnitude
$X$
is equivalent to a physical flux 
\begin{equation}\label{2.2.3a}
{\cal I}_X=10^{(X_{\rm Vega}-X)/2.5}
           \vartheta_0^{({\rm Vega})}{\cal F}_X^{({\rm Vega})}
           {\rm ergs\:sec^{-1}\:cm^{-2}}.
\end{equation}

The monochromatic Eddington flux 
$H_\lambda$
of Vega (Kurucz \cite{kuru1}, 
{\tt fvega.pck}) was integrated
with the same normalized filter functions 
$UBV(RI)_{\rm C}$
of Bessell (\cite{bess1});
$-2.5{\rm lg}{\cal H}_X=$
$-23.835$, $-24.250$, $-23.645$, $-23.083$, $-22.370$
were found for $X=U,B,V,R_{\rm C},I_{\rm C}$ respectively. 
On the other hand the tabular Vega surface magnitudes are
$X_{\rm K}=-19.339$, $-19.755$, $-19.151$, $-18.590$, $-17.875$
if the improved filter functions of Bessell (\cite{bess1}) are used 
(Castelli \cite{cast2}). The difference in these magnitudes is 
$-4.495$ 
i.e. the physical flux of the model is
\begin{equation}\label{2.2.4a}
{\cal F}_X=10^{-(X_{\rm K}-4.495)/2.5}{\rm ergs\:sec^{-1}\:cm^{-2}}
\end{equation}
where
$X_{\rm K}$
is the tabulated magnitude 
$U,B,V,R_{\rm C},I_{\rm C}$ 
(Kurucz \cite{kuru1}).

\end{document}